\documentstyle[12pt]{article}

\def\p{\theta}

\def\pr{\partial }
\def\ca{{\cal A}}
\def\zb {{\bar z}}
\newcommand{\be}{\begin{equation}} \newcommand{\ee}{\end{equation}}
\newcommand{\bea}{\begin{eqnarray}}\newcommand{\eea}{\end{eqnarray}}

\begin{document}
\baselineskip= 24 truept
\begin{titlepage}

\title { Four Dimensional Stringy Black Membrane}


\author{Supriya K. Kar$^1$, S. Pratik Khastgir$^2$ and Gautam Sengupta$^3$
\\
Institute of Physics, Bhubaneswar-751005, INDIA.}

\footnotetext[1]{e-mail:supriya\%iopb@shakti.ernet.in}
\footnotetext[2]{e-mail:pratik\%iopb@shakti.ernet.in}
\footnotetext[3]{e-mail:sengupta\%iopb@shakti.ernet.in}
\date{}
\maketitle

\thispagestyle{empty}

\vskip .7in
\begin{abstract}

\vskip .3in

\noindent An exact conformal field theory describing a four
dimensional singular string background is obtained by chiral gauging
a $U(1)$ subgroup along with translations in $R$ of an $SL(2,R)\times R$
Wess-Zumino-Witten model. It is shown that the target space-time
describes a four dimensional black membrane. Furthermore various
duality transformed solutions are constructed. These are also shown
to correspond to various forms of four dimensional black membranes.

\end{abstract}

\bigskip
\vfil

\rightline{IP/BBSR/92-35}
\rightline{May '92.}
\end{titlepage}

\eject

\noindent In recent years the analogs of singular solutions in
general theory of relativity have come to play an important role in
the understanding of string theories in curved backgrounds.
Intense investigation in this direction has established the existence of
black holes and their higher dimensional generalizations (black
strings and p-branes) as solutions
to the low energy string equations of motion $[1]$. However the exact
conformal field theories describing such singular string backgrounds
were not understood. Very recently the existence of a graviton
dilaton string background describing a two dimensional target space
black hole was established in $[2]$. It was found immediately afterwards
$[3]$ that the exact conformal field theory describing this two dimensional
black hole may be obtained by gauging the axial or the vector
$U(1)$ subgroup of an $SL(2,R)$ Wess-Zumino-Witten (WZW) $[4,5]$ model.
Considerable activity over the past year has unravelled the structure
of this two dimensional black hole and its related generalization
(black strings and p-branes) in string theories $[6,7,8,9,10]$.

In an interesting article $[11]$ Chung and Tye recently showed that
an alternative anomaly free, left right asymmetric, chiral gauging of the
Wess-Zumino-Witten
model was also possible. Recently one of us [SKK]
in collaboration with A.Kumar has shown $[12]$ that the target
space-time of a chiral gauged ${SL(2,R)\over U(1)}$ WZW model corresponds
to a three dimensional black string similar to Horne and Horowitz $[10]$.
In this article we construct an exact conformal field theory
corresponding to a four dimensional black membrane by chiral gauging
the $U(1)$ subgroup along with translations in $R$ of an
$SL(2,R)\times R$ WZW model. We also discuss various duality
transformed solutions exploiting the isometries present. Furthermore
we show that the target space-time structure leads to a
product of a two dimensional black hole $[3]$ and two non compact flat
directions
under certain simplifying assumptions.

The action for the chiral gauged $SL(2,R)$ WZW model is $[11, 12]$
\be
S = S^{WZW} + {k\over 2\pi }\int d^2z\; Tr\; [\; A^{R}_{z}\;
U^{-1}\bar{\pr }U + A^{L}_{\bar z}\; {\pr }U\;
U^{-1} + A^{R}_{z}U^{-1}A^{L}_{\bar z}U\; ] .
\ee
Where $A^R_z\; (A^L_{\bar z})$ in eq.(1) are the $z (\bar
z)$ component of gauge field $A^R_{\mu }\; (A^L_{\mu })$
and $U (z, {\bar z}) \in \ SL(2,R)$.

To obtain the conformal field theory describing a four dimensional
string background we add a single {\it non compact boson} $X$
representing the $R$ direction to the WZW action.
The coupling of the gauge field to the bosonic field $X$
is given by a modification of the gauge invariant action of Isibashi et al
$[9]$. Notice that the complete independence of the left and the
right gauge fields in the chiral gauged model allows us the freedom of
choosing different couplings of the bosonic field $X$ to the gauge field.
We have the following action; $ S_{G}={1\over2\pi}\int d^2z \big[ \pr X
{\bar \pr }X + e_L \ca^{L}_{\zb}
\pr X +e_R\ca^{R}_{z}{\bar \pr}X\big]$.
Adding this action to the WZW action with the gauge field assignments
$A^L_{\bar z} \equiv
\big (-{i\over 2}{\ca }^L_{\bar z}\;\sigma _2\big )$, $A^R_z \equiv
\big (-{i\over 2}{\ca }^R_z\;\sigma _2\big )$ we arrive at the
following action.
\bea
S&&=S_{WZW} + {1\over2\pi}\int\; d^2z \pr X{\bar \pr}X \nonumber\\
&&+ {k\over2\pi}\int\; d^2z \Bigg[ -{1\over2}\ca^{R}_{z}\Big( Tr \big[
\big( \matrix {{0} & {1} \cr {-1} & {0}\cr}\big) U^{-1}{\bar \pr}U\big]
-{2e_R\over k}{\bar \pr}X\Big)\nonumber\\
&&-{1\over2}\ca^{L}_{\bar z}\Big( Tr\big[
\big(\matrix {{0} & {1} \cr {-1} & {0}\cr}\big) \pr U U^{-1}\big]
-{2e_L\over k}\pr X\Big)\nonumber\\
&&+{1\over4}\ca^{R}_{z}\ca^{L}_{\bar z}\Big(Tr \big[ \big( \matrix
{{0} & {1}\cr {-1} & {0}\cr}\big)U^{-1}\big( \matrix {{0} & {1}\cr
{-1} & {0}\cr}\big)U\big] \Big)\Bigg]
\eea

The classical action $S$ in eqn.(2) is invariant under the gauge
transformations $[9,11,12]$
\bea
&&\delta U = v_L(\bar z) U - U v_R(z),\nonumber\\
&&\delta {\ca}^{L}_{\bar z}= - \bar\pr v_L({\bar z})\nonumber\\
&&\delta {\ca}^{R}_{z} = \pr v_R({z}).\nonumber\\
&&\delta X=0.
\eea

\noindent Parametrizing the sigma model field $U$ in terms of the
Euler angles as;

\noindent $ U\equiv exp\; {\big ({i\over 2}\p_L\;\sigma _2\big )}\;
exp\; {\big ({r\over 2}\;\sigma _1\big )}\; exp\; {\big ({i\over 2}\p_R\;
\sigma _2\big )}$ we obtain,
\be
S_{WZW} = {k\over 4\pi } \int d^2z\; \big[-{1\over 2}{\pr }{\p
}_L{\bar\pr }{\p }_L\; -\; {1\over 2}\pr {\p
}_R{\bar\pr }{\p }_R\nonumber\\
+\; {1\over 2}\pr r\bar\pr r -
\cosh r \pr {\p }_R \bar\pr {\p }_L \big],
\ee
\noindent and we have from eqn.(2)
\bea
S =&&S_{WZW}\; + {1\over2\pi}\int \; d^2z\ \pr X {\bar \pr}X \nonumber\\
&& +\; {k\over 4\pi } \int d^2z\; [{\ca }^R_z(\bar\pr
{\p}_R\; +\; \cosh r \bar\pr {\p }_L +{2e_R\over k}{\bar \pr}X)
\nonumber\\
&&+\; {\ca }^L_{\bar
z}(\pr {\p }_L\; +\; \cosh r \pr {\p }_R +{2e_L\over k}\pr X)\;
-\;
{\ca }^L_{\bar z}{\ca }^R_z \cosh r].
\eea
Integrating over the gauge field we obtain the following non linear
sigma model. The
background fields consist of the metric, antisymmetric tensor and a
dilaton field from the one loop correction $[3]$.
\bea
S=&&{k\over 8\pi } \int d^2z\;\big[ \pr r\bar\pr
r+ {\pr }{\p
}_L{\bar\pr }{\p }_L + \pr {\p
}_R{\bar\pr }{\p }_R
+ \big( {4\over k}+{8e_{R}e_{L}\over {k^2 \cosh r}}\big)
\pr X{\bar \pr} X\nonumber\\
&&+{2\over {\cosh r}}\pr {\p }_L \bar\pr {\p }_R
+{4e_R\over {k \cosh r}}\pr \p_{L}{\bar \pr}X
+{4e_L\over {k \cosh r}}\pr
X{\bar \pr} \p_{R}\nonumber\\
&&+{4e_R\over k} \pr \p_{R}{\bar \pr}X + {4e_L\over k}\pr X{\bar \pr}\p_{L}
+R^{(2)}(\ln \cosh r + c)\big]
\eea
Where the background metric ${(G_{\mu\nu })}$ and
antisymmetric tensor ${(B_{\mu\nu })}$ are given as follows
\be
G =\left (\matrix {{1} & {0} & {0} & {0}\cr {0} & {1} & {{1\over
{\cosh r}}} & {{2\over k}(e_L+{e_R\over \cosh r})}\cr {0}
& {{1\over \cosh r}} & {1} & {{2\over k}(e_R+{e_L\over \cosh r})}\cr
{0} & {{2\over k}(e_L+{e_R\over \cosh r})} &
{{2\over k}(e_R+{e_L\over \cosh r})}
& {({4\over k}+{8e_{R}e_{L}\over {k^2 \cosh r}})}\cr }\right )
\ee
\be
B =\left (\matrix {{0} & {0} & {0} & {0}\cr {0} & {0} & {{1\over
{\cosh r}}} & {-{2\over k}(e_L-{e_R\over \cosh r})}\cr {0} &
{-{1\over \cosh r}} & {0} & {{2\over k}(e_R-{e_L\over \cosh r})}\cr
{0} & {{2\over k}(e_L-{e_R\over \cosh r})} &
{-{2\over k}(e_R-{e_L\over \cosh r})}
& {0}\cr }\right ),
\ee
where we have modded out the constant ${k\over4}$.
The one loop contribution to dilaton is $\phi = - \ln \cosh r + const.$

Observe that the background fields depend on a single coordinate ($r$ in
our case). Such backgrounds have been
extensively discussed by Meissner and Veneziano $[13]$ and the field
equations satisfied by these background fields to one loop in the
sigma model has been obtained. We have explicitly verified that
the background configuration obtained by us satisfies the
complete set of field equations given in ref.$[13]$ for the
cosmological term $V=1$.
Thus the non linear sigma model
in (6) describes a string background with scalar curvature
$R = -{7\over 2}\; {1\over {\cosh^2r}}$.
This shows that the target space-time contains a curvature singularity. However
it does not occur for any real value of $r$ in the coordinate system
of our choice. Applying an orthogonal transformation $[G\rightarrow\
O^T G O]$ through the matrix
\be
O=\left (\matrix {{1} & {0} & {0} & {0}\cr {0} & {{1\over {\sqrt 2}}}
& {-{1\over {\sqrt 2}}} & {0}\cr
{0} & {{1\over {\sqrt 2}}} & {{1\over {\sqrt 2}}} & {0}\cr
{0} & {0} & {0} & {1}\cr
}\right )\; ,
\ee
to simplify the form of the target space-time we arrive at the
following expressions for $G$ and $B$ such that,
\be
G=\left (\matrix {{1} & {0} & {0} & {0}\cr {0} &
{(1+{1\over
\cosh r})} & {0} & {{{\sqrt 2}\over k}e_+\big(1+{1\over
\cosh r}\big)}\cr
{0} & {0} & {\big(1-{1\over
\cosh r}\big)} & {{{\sqrt 2}\over k}e_-\big(1-{1\over
\cosh r}\big)}\cr
{0} & {{{\sqrt 2}\over k}e_+\big(1+{1\over
\cosh r}\big)} & {{{\sqrt 2}\over k}e_-\big(1-{1\over
\cosh r}\big)} & {\big({4\over k}+{8e_{R}e_{L}\over {k^2\cosh r}}\big)}\cr
}\right )\;\; ,
\ee
\be
B=\left (\matrix {{0} & {0} & {0} & {0}\cr {0} & {0} & {{1\over
\cosh r}} & {{{\sqrt 2}\over k}e_-\big(1+{1\over
\cosh r}\big)}\cr
{0} & {-{1\over \cosh r}} & {0} & {{{\sqrt 2}\over k}e_+\big(1-{1\over
\cosh r}\big)}\cr
{0} & {-{{\sqrt 2}\over k}e_-\big(1+{1\over
\cosh r}\big)} & {-{{\sqrt 2}\over k}e_+\big(1-{1\over
\cosh r}\big)} & {0}\cr
}\right )\; ,
\ee
where $e_R+e_L=e_+$ and $e_R-e_L=e_-$.
The invariant line element reduces to
\bea
ds^2&&=dr^2+\Big( 1+{1\over \cosh r}\Big)dx^2 +\Big( 1-{1\over
\cosh r}\Big)dy^2 +\Big({4\over k}+{8e_{R}e_{L}\over {k^2 \cosh r}}\Big)dz^2
\nonumber\\
&&+{2{\sqrt 2}\over k}e_+\Big( 1+{1\over \cosh r}\Big)dxdz
+{2{\sqrt 2}\over k}e_-\Big( 1-{1\over \cosh r}\Big)dydz
\eea
where $r,\;x\;,y\;,z$ are the new space-time coordinate.
 From the form of the line element (12) we observe that the target space-time
described by the chiral gauged ${SL(2,R)\times R\over U(1)}$ WZW
model is asymptotically flat. In the limit the WZW level
$k\rightarrow\ \infty$ this reduces to the three dimensional black
string of Horne and Horowitz $[10, 12]$. In the extremal limit of not
gauging $R$ at all $i.e$ ${\it e_R=e_L=0}$ the target space-time
describes a product of the three dimensional black string $[10,12]$ and a
noncompact flat direction. So the metric (12) describes
a four dimensional black membrane with axionic charge.

We now apply the duality transformations given in Nojiri $[14]$ exploiting the
isometry
in the $x$-direction. \def\tg {{\tilde G}}  \def\tb {{\tilde B}}
The dual configuration is
\be
\tg=\left (\matrix {{1} & {0} & {0} & {0}\cr {0} & {{\cosh
r\over {\cosh r+1}}} & {-{1\over {\cosh r+1}}} & {-{{\sqrt 2}\over k}e_-}
\cr
{0} & {-{1\over {\cosh r+1}}} & {{\cosh r\over {\cosh r+1}}} &
{{{\sqrt 2}\over k}e_-}\cr
{0} & {-{{\sqrt 2}\over k}e_-} & {{{\sqrt 2}\over k}e_-} &
{\big({4\over k}-{8e_{R}e_{L}\over {k^2}}\big)}\cr
}\right )\; ,
\ee
the antisymmetric tensor is reduced to a constant value
\be
\tb=\left (\matrix {{0} & {0} & {0} & {0}\cr {0} & {0} & {0} &
{-{{\sqrt 2}\over k}e_+}\cr
{0} & {0} & {0} & {{{\sqrt 2}\over k}e_+}\cr
{0} & {{{\sqrt 2}\over k}e_+} & {-{{\sqrt 2}\over k}e_+} & {0}\cr
}\right )\;\; ,
\ee
and $ {\tilde \phi }=-\ln (\cosh r+1) + const.=-\ln \cosh^2 {r\over2} + const.$

It is established in Rocek et al $[15]$ that the
dual manifold is described by the same conformal field theory
when the isometry under consideration is compact. However if the
isometry is along a non compact direction then the dual manifold is actually
an orbifold obtained by modding out the translations in that
direction. For the conformal field theory this implies that the momentum modes
are replaced by the winding modes $[15]$. For our case the dual
solution must be admitted with the foregoing caveat as the isometries
considered are along two non compact directions.

Once more we apply an orthogoanl transformation to bring the
expressions to a familiar form using the matrix $O$ in eqn.(9)
to arrive at
\be
{\hat G}=\left (\matrix {{1} & {0} & {0} & {0}\cr {0} &
{\tanh^2 ({r\over2})} & {0} & {0}\cr
{0} & {0} & {1} & {{2\over k}e_-}\cr
{0} & {0} & {{2\over k}e_-} & {({4\over k}-{8e_{R}e_{L}\over {k^2}})}\cr
}\right )\; ,
\ee
\be
{\hat B}=\;\left (\matrix {{0} & {0} & {0} & {0}\cr {0} &
{0} & {0} & {0}\cr
{0} & {0} & {0} & {{2\over k}e_+}\cr
{0} & {0} & {-{2\over k}e_+} & {0}\cr
}\right )\; ,
\ee
and $\phi $ remains unchanged.

The invariant line element is now
\be
ds^2=dr^2 +\tanh^2({r\over2})dx^2 +dy^2 +{4\over k}e_-dydz + dz^2
\ee
where we have rescaled $z={\sqrt {({4\over k}-{8e_{R}e_{L}\over
k^2})}}\; z$.

We now proceed to discuss certain interesting limiting cases of our solution.
Earlier we have stated that the complete independence of the gauge
fields in the left and the right sectors allows us the freedom of
choosing different couplings to the non compact boson $X$.
This corresponds to the most general description of the target space-time
geometry.
We observe that with the identifications $e_R=e_L=e$ or $e_R=-e_L=e$ we
arrive at simplified form of the target space-time. Observe from eqn.(17)
that for the case $e_R=e_L=e$ our solution corresponds
to the product of the two dimensional black hole $[3]$ and two non compact
flat directions. For the case $e_R=-e_L=e$ we obtain a target space-time
with zero torsion field and non diagonal form of the metric. Notice that
the presence of two non compact isometries in the $x$ and the $y$-direction
leads to two differing descriptions of the target space time using
the duality transformations corresponding to these isometries. It is
possible to show that using the $y$ isometry for the case $e_R=-e_L$
we arrive at the same target space time for $e_R=e_L$. So these two
solutions are in one sense dual to each other.

On the other hand if we had employed the duality transformations
corresponding to the $y$ isometry in eqns.(10,11) we would have
arrived at a dual description of the target space-time described in eqn.(17).
In this case also we can show that the two different identifications of the
gauge couplings as above lead to two different target space-time which are
related by the duality corresponding to the residual $x$ isometry as opposed
to the $y$ isometry in the previous case.

To conclude we have obtained an exact conformal field theory
describing a novel space-time geometry corresponding to
stationary black membrane in four dimensions. This is represented by a
chiral gauged ${SL(2,R)\times R\over U(1)}$ WZW
model. We discuss several interesting aspects of our solution
corresponding to the identifications of the left and the right gauge
couplings of the gauge fields to the non compact bosonic field $X$.
Furthermore we exploit the two different isometries present
to implement duality transformations and arrive at differing dual
descriptions of the target space-time geometry describing a black
membrane in four dimensions. We mention {\it en passant}
that by choosing the boson $X$ as compact a space-time
$U(1)\times U(1)$ gauge background along with a scalar field is induced
in addition to the metric, antisymmetric tensor
field and the dilaton. We have verified that this leads to a three dimensional
electrically charged black string solution analogous to that obtained in Horne
and Horowitz $[10]$. The solution obtained by us may be generated
from the three dimensional solution in $[12]$ by application of the
algorithm outlined in $[16]$ for generating classical solution in higher
dimensions starting from a lower dimensional solution. This
clearly illustrates that it is possible to arrive at our solution starting
from a suitable string effective action $[13]$ with the cosmological
term $V=1$ in four dimensional space-time.
It would be interesting to investigate the thermodynamic and the
cosmological implications of our solution.

\bigskip
\bigskip

\noindent {\bf Acknowledgements:}\hfil\break
All of us would like to thank Alok Kumar and Jnanadeva Maharana for
suggestions and many interesting discussions.

\vfil
\eject

\def\npb{Nucl.Phys.\ {\bf B}}
\def\plb{Phys. Lett.\ {\bf B}}
\def\prl{Phys. Rev. Lett.}

\baselineskip 12pt
\vfil
\eject

\end{document}